# Self-navigated 3D diffusion MRI using an optimized CAIPI sampling and structured low-rank reconstruction

Ziyu Li, Karla L. Miller, Xi Chen, Mark Chiew, Wenchuan Wu

**Abstract—** **3D multi-slab acquisitions are an appealing approach for diffusion MRI because they are compatible with the imaging regime delivering optimal SNR efficiency. In conventional 3D multi-slab imaging, shot-to-shot phase variations caused by motion pose challenges due to the use of multi-shot k-space acquisition. Navigator acquisition after each imaging echo is typically employed to correct phase variations, which prolongs scan time and increases the specific absorption rate (SAR). The aim of this study is to develop a highly efficient, self-navigated method to correct for phase variations in 3D multi-slab diffusion MRI without explicitly acquiring navigators. The sampling of each shot is carefully designed to intersect with the central kz=0 plane of each slab, and the multi-shot sampling is optimized for self-navigation performance while retaining decent reconstruction quality. The kz=0 intersections from all shots are jointly used to reconstruct a 2D phase map for each shot using a structured low-rank constrained reconstruction that leverages the redundancy in shot and coil dimensions. The phase maps are used to eliminate the shot-to-shot phase inconsistency in the final 3D multi-shot reconstruction. We demonstrate the method's efficacy using retrospective simulations and prospectively acquired in-vivo experiments at 1.22 mm and 1.09 mm isotropic resolutions. Compared to conventional navigated 3D multi-slab imaging, the proposed self-navigated method achieves comparable image quality while shortening the scan time by 31.7% and improving the SNR efficiency by 15.5%. The proposed method produces comparable quality of DTI and white matter tractography to conventional navigated 3D multi-slab acquisition with a much shorter scan time.**

## I. INTRODUCTION

High-resolution diffusion MRI can provide detailed information about tissue microstructure and accurate representation of intricate fiber arrangements [1, 2]. However, the effectiveness of diffusion MRI is limited by its inherently low SNR, which further decreases as the resolution increases.

3D multi-slab is a promising approach for high-resolution diffusion MRI due to its ability to achieve short TR=1-2s [3-6] for optimal SNR efficiency for spin-echo based diffusion MRI. This technique divides the entire imaging volume into multiple thin slabs with 10-20 slices per slab, typically employing a 3D multi-shot echo-planar imaging (EPI) trajectory for high efficiency [3-6]. However, this method is sensitive to motion-induced shot-to-shot phase variations that degrade image quality when combining data from different shots [7, 8].

Conventional methods to correct the phase inconsistency in multi-shot diffusion MRI require navigators [3-8]. The navigators are incorporated into a model-based reconstruction to correct phase errors [7]. Previous studies have shown that the motion-induced phase within each 3D slab can be approximated by a 2D navigator if the slabs are sufficiently thin (i.e., <2 cm) [3-6]. This 2D navigator necessitates an additional spin echo, leading to an extended TR and increasing the scan time by 25%-50% [5, 9]. Furthermore, the inclusion of another RF refocusing pulse increases the specific absorption rate (SAR).

Several studies have investigated the feasibility of navigator-free 2D multi-shot diffusion imaging. Multiplexed sensitivity-encoding (MUSE) reconstructs a 2D phase map from each under-sampled shot [10]. Despite the relatively high under-sampling factor per shot, clean phase maps can be reconstructed by exploiting the smoothness of motion-induced phase. Methods avoiding an explicit phase map have been proposed [11, 12], which leverage the magnitude consistency between shots to create a structured Hankel matrix and estimate the missing data in each shot using structured low-rank (SLR) matrix completion [13].

Extending these methods for navigator-free 3D multi-slab diffusion imaging is intrinsically challenging. Specifically, to extend MUSE to 3D requires estimating a 2D phase map from each shot to capture in-plane phase variations. However, for conventional 3D Cartesian EPI (e.g., [3-5]), shots covering peripheral kz planes encode high spatial frequency information, and estimating in-plane smooth phase variations is challenging. Additionally, in conventional 3D multi-slab diffusion imaging, each shot is effectively subjected to a high under-sampling factor with little redundancy between shots, which significantly

W.W. is supported by the Royal Academy of Engineering (RF\201819\18\92). K.L.M. is supported by the Wellcome Trust (WT202788/Z/16/A). M.C. is supported by the Canada Research Chair Program. This study is supported by the NIHR Oxford Health Biomedical Research Centre (NIHR203316). The views expressed are those of the author(s) and not necessarily those of the NIHR or the Department of Health and Social Care. The Wellcome Centre for Integrative Neuroimaging is supported by core funding from the Wellcome Trust (203139/Z/16/Z and 203139/A/16/Z).

Z.L., K.L.M., X.C, M.C., W.W. are with Wellcome Centre for Integrative Neuroimaging, FMRIB, Nuffield Department of Clinical Neurosciences, University of Oxford, Oxford, UK (correspondence to: wenchuan.wu@ndcn.ox.ac.uk (W.W.)).

X.C. is also with Department of Radiological Sciences, David Geffen School of Medicine at UCLA, Los Angeles, California, USA. M.C. is also with Physical Sciences, Sunnybrook Research Institute and Department of Medical Biophysics, University of Toronto, Toronto, Canada.

hampers the feasibility of extending 2D SLR approach to 3D. Therefore, simplistically applying the 2D SLR approach to 3D multi-slab diffusion imaging might be impractical.

In this study, we present a novel acquisition and reconstruction framework for self-navigated 3D multi-slab diffusion MRI. We propose a new 3D EPI sampling pattern that enables self-navigation with minimized k-space gaps in the shot-combined sampling and minimized overlapping between shots. A k-space based SLR constrained reconstruction is leveraged to jointly exploit data redundancy across shots and coils to reconstruct high-quality phase maps. Our multi-shot reconstruction incorporates phase error correction to calculate the final image. The proposed method's efficacy is evaluated through in-vivo experiments on a 7T scanner. The results demonstrate comparable image quality and significantly improved SNR efficiency compared to navigator-based methods. The resulting whole-brain diffusion tensor imaging (DTI) and tractography results highlight our proposed method's potential to enable high-resolution 3D multi-slab diffusion imaging with improved time efficiency.

## II. THEORY

### A. Review of 3D multi-slab diffusion imaging

3D multi-slab diffusion imaging uses multi-shot EPI to encode a 3D k-space for each slab. The acquired k-space signal for the $j^{th}$ shot can be represented as:

$$y_j(\mathbf{k}) = \int \rho_j(\mathbf{r})\phi_{d_j}(\mathbf{r})e^{-i2\pi\mathbf{k}^T\mathbf{r}}d\mathbf{r} + n_j(\mathbf{k}), \qquad (1)$$

where $y_j$ is the multi-coil diffusion k-space data of shot $j$, $\rho_j$ is the multi-coil phase error-free image signal for the $j^{th}$ shot, $\phi_{d_j} = e^{i\psi_j}$, and $\psi_j$ is the non-diffusive motion induced phase in shot $j$, $j \in [0, N_{shot} - 1]$, where $N_{shot}$ denotes the total numbers of shots. In conventional 3D multi-slab diffusion imaging, each shot covers a kz plane using a single EPI readout for efficient data acquisition (Fig. 1a), and $N_{shot}$ equals to the number of encoded $k_z$ planes ($N_{k_z}$). $n_j$ is the additive Gaussian noise in shot $j$. This signal formulation described in Eq. 1 can be represented in a matrix form as:

$$y_j = \mathcal{A}_j(X) + n_j = \mathcal{M}_j F \phi_{d_j} X + n_j, \qquad (2)$$

where $\mathcal{A}_j$ is the forward model of shot $j$, $X$ is the multi-coil phase error-free 3D image volume, $\mathcal{M}_j$ is the k-space sampling matrix of shot $j$, and $F$ is the 3D Fourier Transform.

Non-diffusive motions during diffusion encoding introduced shot-dependent phase errors $\phi_d$ can lead to significant image corruptions if not corrected. Therefore, accurate information of phase errors $\phi_{d_j}$ is necessary for the reconstruction of $X$.

In conventional 3D multi-slab diffusion imaging, a 2D navigator is acquired at $k_z = 0$ for each shot using an extra refocusing RF pulse (i.e., the navigator samples the secondary spin echo) and the phase of the navigator is used as an estimation of motion induced phase errors (Fig. 1a). However, navigator acquisition suffers from several drawbacks, including prolonged scan time, increased SAR associated with a second spin echo, and reduced SNR efficiency.

### B. Review of SLR reconstruction for 2D Navigator-Free diffusion imaging

The 2D SLR reconstruction (e.g., MUSSELS [11, 12, 14]) leverages the data redundancy across different shots to effectively restore missing data. Consequently, it enables accurate estimation of magnitude and phase information for each shot without the need for explicitly acquired phase maps.

MUSSELS assumes different shots share the same underlying magnitude image $\mathbf{m}$ despite their different phases:

$$\mathbf{m} = x_i \Phi_i^H = x_j \Phi_j^H, \qquad i, j \in [0, N_{shot} - 1], \qquad (3)$$

where $x_i$ and $x_j$ are the complex image for shot $i$ and $j$, respectively, $\Phi = \phi_d\phi_c$ includes the diffusion-related phase $\phi_d$ and coil-related phase $\phi_c$, and $\Phi^H$ denotes the conjugate of $\Phi$. This leads to the establishment of an annihilation relation in both the image domain (Eq. 4) and the Fourier domain (Eq. 5):

$$x_i\Phi_j - x_j\Phi_i = 0, \qquad i, j \in [0, N_{shot} - 1], \qquad (4)$$

$$\hat{x}_i * \widehat{\Phi}_j - \hat{x}_j * \widehat{\Phi}_i = 0, \qquad i, j \in [0, N_{shot} - 1], \qquad (5)$$

where $\hat{x}$ and $\widehat{\Phi}$ denotes the Fourier Transform of $x$ and $\Phi$, respectively, and $*$ denotes the convolution operation. As $\Phi$ is typically smooth in image space, $\widehat{\Phi}$ should be support limited in the Fourier domain. Utilizing the block-Hankel matrix formulation $H_1(\hat{x})$ detailed in previous work [11], Eq. 5 can be expressed in a matrix form:

$$\begin{bmatrix} H_1(\hat{x}_i) & H_1(\hat{x}_j) \end{bmatrix} \begin{bmatrix} vec(\widehat{\Phi}_j) \\ -vec(\widehat{\Phi}_i) \end{bmatrix} = 0, \qquad i, j \in [0, N_{shot} - 1]. \quad (6)$$

This relation holds for all pairs of shots as they are all assumed to share the same magnitude image, Thus, the structured matrix

$$H(\hat{x}) = \begin{bmatrix} H_1(\hat{x}_0) & H_1(\hat{x}_1) & ... & H_1(\hat{x}_{N_{shot}-1}) \end{bmatrix} \qquad (7)$$

exhibits a low-rank property due to the existence of a non-trivial null space P ($H(\hat{x})P = \mathbf{0}$):

$$P = \begin{bmatrix} \widehat{\Phi}_1 & 0 & 0 & \widehat{\Phi}_2 \\ -\widehat{\Phi}_0 & \widehat{\Phi}_2 & 0 & 0 \\ 0 & -\widehat{\Phi}_0 & ... & 0 & -\widehat{\Phi}_0 & ... \\ \vdots & \vdots & & \vdots & \vdots \\ 0 & 0 & & \widehat{\Phi}_{N_{shot}-1} & 0 \\ 0 & 0 & & -\widehat{\Phi}_{N_{shot}-2} & 0 \end{bmatrix}. \qquad (8)$$

The 2D SLR reconstruction enforces the low-rankness of $H(\hat{x})$ by minimizing $\left\| H(\hat{x}) \right\|_*$ during iterative reconstruction to facilitate high-fidelity navigator-free 2D diffusion imaging.

However, the extension of 2D SLR reconstruction to 3D multi-slab diffusion imaging presents notable challenges. While previous approaches, such as gSlider [15, 16], have effectively applied SLR to high-resolution multi-shot diffusion imaging, they essentially reconstructed RF-encoded 2D k-space. Our work focuses on leveraging SLR for 3D k-space reconstruction. A direct extension to 3D SLR (by substituting 2D images in Eq. 3 with 3D volumes) requires reconstructing an entire 3D volume from a single shot, which undergoes exceptionally high under-sampling factor for 3D multi-slab imaging. For instance, if a single slab is encoded with 10 shots and an acceleration factor of $R_y = 3$ is applied along the phase-encoding direction, the under-sampling factor for each shot would amount to 30. Furthermore, the construction of Hankel matrices using 3D volumes in Eq. 7 would entail substantial computational costs and feasibility concerns.

### C. Self-navigated 3D multi-slab imaging framework

Our method, illustrated in Fig. 1b, aims to integrate the SLR reconstruction into 3D multi-slab diffusion imaging to eliminate the need for navigator acquisition. This is achieved with an extended CAIPI [17] sampling trajectory where each shot covers a wide range of kz planes and intersects with the central kz=0 plane (the intersections are called "self-navigation points" hereafter). The self-navigation points are used to reconstruct a fully-sampled kz=0 plane for each shot, akin to a 2D navigator but without requiring additional scan time. The sampling trajectories of all shots are jointly optimized for robust self-navigation and multi-shot reconstruction. The limited self-navigation points result in high under-sampling. We developed a SPIRiT-based [18] SLR reconstruction that exploits the redundancy between coils and shots to produce a robust estimation of motion-induced phase error for each shot. This is then incorporated into the multi-shot reconstruction to eliminate shot-to-shot phase inconsistencies as in conventional 3D multi-slab diffusion imaging reconstruction [3-7] (Fig. 1).

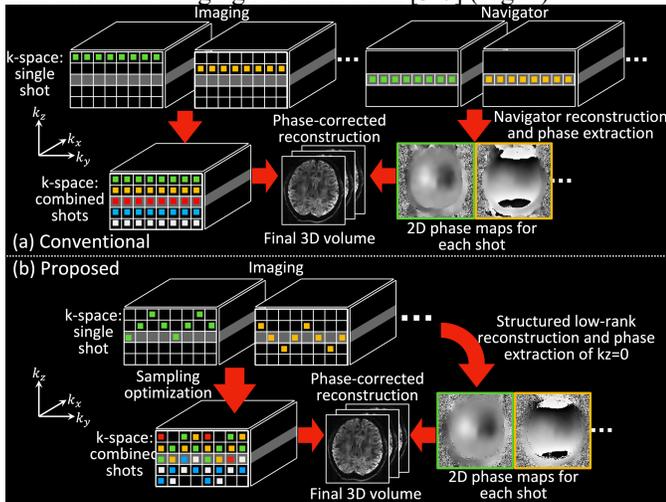

Fig. 1. Comparison between conventional navigated 3D multi-slab imaging (a) and the proposed self-navigated imaging framework (b). In the conventional approach (a), phase information is obtained from a separately acquired 2D navigator. In contrast, the proposed method (b) extracts shot phase directly from the imaging data using a novel acquisition and reconstruction framework, eliminating the need for acquiring a separate navigator.

### D. K-space sampling optimization

The sampling coverage of each shot's trajectory is expanded along kz by inserting kz blips between readout gradients, such that each shot traverses through the central kz=0 plane to provide the self-navigation points. The kz width of each shot $w$ was set to $w = \text{floor}(N_{kz}/2) + 1$, the minimal value to ensure each shot intersects with the central kz=0 plane. A fundamental sampling configuration was established for an individual shot with a fixed ky period and kz width (Fig. 2a), based on which the sampling pattern was designed.

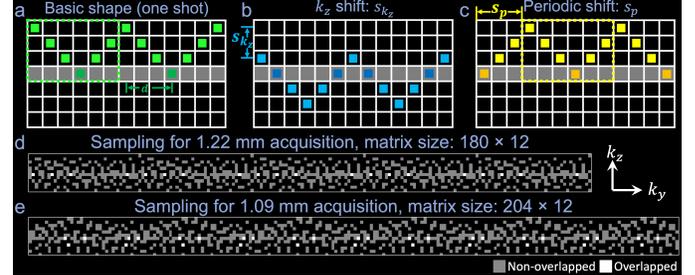

Fig. 2. K-space sampling optimization. The basic shape of one shot of the proposed extended CAIPI sampling (a), the illustration of sampling parameters: kz shift $s_{k_z}$(b) and periodic shift $s_p$ (c). The acceleration along the phase encoding direction (ky) is not illustrated. The intersections with central kz=0 plane (self-navigation points) are marked in dark green (a), dark blue (b), and orange (c). The parameter $d$ which denotes the shortest distance between self-navigation points and the ky-kz center is also illustrated in (a). The dashed green (a) and yellow (c) boxes indicate one period of the periodic sampling. The resulting sampling patterns for 1.22 mm acquisition (matrix size: 180×12, d) and 1.09 mm acquisition (matrix size: 204×12, e) using the proposed sampling optimization are demonstrated. The non-overlapped and overlapped points are marked in gray and white, respectively.

Drawing from this basic sampling, each shot could be acquired with a different pattern by modifying several parameters characterizing the sampling. The kz shift $s_{k_z}$ can be altered to determine the sampling's vertical position (Fig. 2b) ($s_{k_z} \in [0, N_{k_z} - w]$). The periodic shift $s_p$ controls the starting point of the periodic sampling (Fig. 2c), ($s_p \in [0, 2w - 3]$). Acceleration is usually applied along the phase encoding direction (ky) to shorten TE and mitigate geometric distortion (not illustrated in Fig. 1 and Fig. 2). Therefore, the sampling pattern can also be modified with a ky shift $s_{k_y}$ ($s_{k_y} \in [0, 1, .., R_y - 1]$, where $R_y$ is the acceleration factor along ky).

The sampling pattern can impact the image reconstruction quality in various ways. First, distinct shots might overlap, reducing sampling efficiency. Second, large gaps in k-space coverage could introduce substantial artifacts in the reconstructed images [19]. Lastly, the positioning of self-navigation points also affects phase correction performance. Ideally, each shot should incorporate some self-navigation points positioned close to the ky-kz center to capture sufficient low-frequency data for precise 2D phase map reconstruction. An optimal sampling scheme should strive to minimize overlap and gaps in k-space coverage, while ensuring each shot includes some self-navigation points near the ky-kz center.

We developed a framework to optimize the sampling pattern through metrics of overlap, gaps and self-navigation points. The metric $o_i$ is the number of overlap points in the $i$-shot-combined sampling:

$$o_i = T_o \left( \sum_{j=0}^{i-1} M_j(s_{k_{y_j}}, s_{k_{z_j}}, s_{p_j}) \right), \quad (9)$$

where $M_j(s_{k_{y_j}}, s_{k_{z_j}}, s_{p_j})$ is the sampling mask for shot $j$, specified by parameters $s_{k_{y_j}}, s_{k_{z_j}}, s_{p_j}$, respectively. $T_o$ denotes an operation counting the number of elements greater than one within the combined sampling mask $\sum_{j=0}^{i-1} M_j$.

The metric $g_i$ reflects the number of gaps larger than 3×3 in the $i$-shot-combined sampling:

$$g_i = T_g\left(\left(\sum_{j=0}^{i-1} M_j(s_{k_{y_j}}, s_{k_{z_j}}, s_{p_j})\right) * \begin{bmatrix} 1 & 1 & 1 \\ 1 & 1 & 1 \\ 1 & 1 & 1 \end{bmatrix}\right), \quad (10)$$

where $T_g$ counts the number of zeros in the convolution result.

The metric $d$ is the shortest distance between self-navigation points and the ky-kz center for each shot (Fig. 2a). It is essential for each shot to have a small value of $d$ to provide low-frequency information for accurate reconstruction of 2D phase maps. We establish a maximum allowable distance $d_{max}$.

We optimize the sampling by solving the following problem:

$$\arg \min_{s_{k_{y_i}}, s_{k_{z_i}}, s_{p_i}} o_i + g_i + d_i, \qquad s.t., \qquad d_i \leq d_{max}. \quad (11)$$

The optimization is solved efficiently using greedy search in a shot-by-shot manner. Importantly, the first shot is designed to traverse the kz=0 plane without kz blip to accurately capture the magnitude information of kz=0, which benefits the SLR reconstruction of the phase maps (see Section E).

### E. SPIRiT-based SLR reconstruction for phase map estimation

We address the extreme under-sampling of the phase maps by jointly leveraging the shared information across shots and coils using a SPIRiT [18, 20]-based SLR reconstruction.

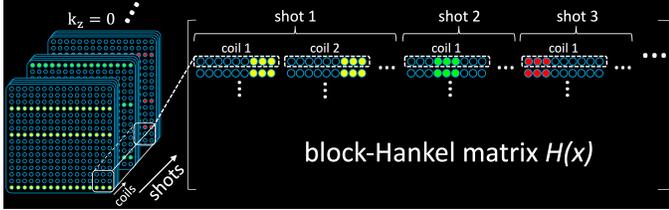

Fig. 3. Formulation of the block-Hankel matrix. The block-Hankel matrix for the proposed SPIRiT-based structured low-rank reconstruction leverages the low-rank properties in both coil and shot dimensions.

The careful design of the sampling facilitates the adaptation of 2D SLR method for the reconstruction of the central kz=0 plane for each shot. Equations 3-8 can be readily applied to construct a low-rank block-Hankel matrix for kz=0 data, assuming $x_j$ represents the kz=0 image for shot $j$. Notably, unlike the original 2D SLR formulation [11, 12, 14] where $x_j$ is the coil-combined image, we utilize the multi-coil data to construct the block-Hankel matrix. Because coil sensitivity is smooth with limited k-space support, the multi-coil formulation promotes the low-rankness of the block-Hankel matrices [21]. The formulation of $H(\hat{x})$ is illustrated in Fig. 3. By minimizing $||H(\hat{x})||_*$, we enforce low-rank and harness the redundancy in both the shot and coil dimensions. This constraint promotes similarity among the magnitude images derived from different shots. The use of one shot that traverses the kz=0 plane without kz blips provides accurate magnitude information $\rho'$ to improve the robustness and accelerate the convergence.

Our SPIRiT-based SLR reconstruction recovers the 2D phase maps for each shot from the self-navigation points by solving:

$$\arg \min_{\hat{x}} ||\mathcal{M}_0\hat{x} - y_0||_2^2 + \lambda_1 ||(G_0 - I)\hat{x}||_2^2 + \lambda_2 ||H(\hat{x})||_*, (12)$$

where $\mathcal{M}_0$ selects self-navigation points, $\hat{x}$ is the fully-sampled k-space data for different shots at kz=0, $y_0$ is the acquired self-navigation points, $G_0$ is the SPIRiT kernel trained on the kz=0 calibration data (see below), $I$ is the identity matrix, and $\lambda_1, \lambda_2$ are the hyperparameters for the SPIRiT and SLR constraints.

The SPIRiT kernel $G_0$ is trained on calibration data similarly to GRAPPA [22] and performs convolutions in k-space to reach self-consistency:

$$\hat{x} = G_0\hat{x}, \quad (13)$$

which is achieved by minimizing $||(G_0 - I)\hat{x}||_2^2$ in Eq. 12.

We reformulate Eq. 12 to solve it iteratively using alternating direction method of multipliers (ADMM) [23]:

$$\arg \min_{\hat{x}} ||\mathcal{M}_0\hat{x} - y_0||_2^2 + \lambda_1 ||(G_0 - I)\hat{x}||_2^2 + \lambda_2 ||z||_*.$$
$$s.t., \quad z - H(\hat{x}) = 0, \quad (14)$$

In the $k^{th}$ iteration, Eq. 14 is further split into the following three subproblems:

$$\hat{x}^k = \arg \min_{\hat{x}} ||\mathcal{M}_0\hat{x} - y_0||_2^2 + \lambda_1 ||(G_0 - I)\hat{x}||_2^2 + \left(\frac{\beta}{2}\right) ||z^{k-1} - H(\hat{x}) + u^{k-1}||_2^2, \quad (15)$$

$$z^k = \arg \min_z \lambda_2 ||z||_* + \left(\frac{\beta}{2}\right) ||z - H(\hat{x}^k) + u^{k-1}||_2^2, \quad (16)$$

$$u^k = u^{k-1} + z^k - H(\hat{x}^k), \quad (17)$$

where $\beta$ is the dual update step length. As all shots share the same magnitude information, an explicit magnitude image $\rho'$ for kz=0 is obtained from the kz=0 traversing shot, which is included in Eq. 15 as another constraint:

$$\hat{x}^k = \arg \min_{\hat{x}} ||\mathcal{M}_0\hat{x} - y_0||_2^2 + \lambda_1 ||(G_0 - I)\hat{x}||_2^2 + \lambda_3 ||\hat{x} - \rho'\Phi^{k-1}||_2^2 + \left(\frac{\beta}{2}\right) ||z^{k-1} - H(\hat{x}) + u^{k-1}||_2^2, \quad (18)$$

where $\Phi^{k-1}$ is the image phase of $\hat{x}^{k-1}$. In practice, the constraint $||\hat{x} - \rho'\Phi^{k-1}||_2^2$ was found effective in improving the robustness and convergence speed of the reconstruction. We solve Eq. 16 by applying singular value decomposition (SVD) and hard thresholding similar to previous work [11, 21, 24]. Eq. 18 is solved using the conjugate gradient (CG) method.

A 2D phase map can then be extracted from the reconstructed kz=0 image $\hat{x}_i$ for each shot:

$$\phi_{d_i} = \frac{S^H F^{-1} \hat{x}_i}{\left|\left|S^H F^{-1} \hat{x}_i\right|\right|}, i \in [0, N_{shot} - 1], \qquad (19)$$

where $S$ is the coil sensitivity maps for kz=0, which can be obtained from kz=0 calibration data. According to Eq. 2, the 2D phase maps can be incorporated into the forward model to address the phase variations in the multi-shot reconstruction:

$$X = \arg\min_X \sum_i \left|\left|\mathcal{M}_i F \phi_{d_i} F^{-1} X - y_i\right|\right|_2^2$$
$$+ \lambda_4 \left|\left|(G_{slab} - I)X\right|\right|_2^2, \qquad (20)$$

where $X$ is the desired phase error-corrected 3D slab image, $G_{slab}$ is the SPIRiT kernel trained on coil calibration data for the whole slab, and $\lambda_4$ is the SPIRiT regularization weight of the final phase-corrected reconstruction.

## III. METHODS

### A. Data acquisition

A 3D multi-slab spin-echo diffusion MRI sequence [4] was modified to integrate the proposed CAIPI sampling for self-navigation (referred to as "Self-nav CAIPI" hereinafter).We used $N_{kz}$=12, $R_y$=3, $d_{max}$=15, such that, $s_{kz} \in [0,5]$, $s_{k_y} \in [0,2]$, $s_p \in [0,11]$. For the $i^{th}$ shot, we exhaustively explored all combinations of $s_{k_{y_i}}, s_{kz_i}, s_{p_i}$ to identify lowest value for the cost function $o_i + g_i + d_i$. The resulting sampling patterns for 1.22 mm acquisition (matrix size: 180×12 for ky-kz plane) and 1.09 mm acquisition (matrix size: 204×12 for ky-kz plane) are demonstrated in Fig. 2d and Fig. 2e. There is little overlapping and no substantial k-space gaps. The samplings are denser near k-space center with an overall random distribution.

Some experiments included an explicit navigator acquisition with 64 phase encoding lines and matching the imaging echo for phase encode direction, Ry, and echo spacing (ES).

Subjects were scanned on a Siemens 7T scanner (Siemens Magnetom, Erlangen, Germany) using a 32-channel receive coil. Written informed consent in accordance with local ethics was obtained from each subject.

Table I summarizes six experimental protocols. Images were acquired at 1.22 and 1.09 mm isotropic resolutions (180×180 and 204×204 matrix, ES=0.76 and 0.82 ms, respectively). Unless otherwise specified, acquisitions used 13 slabs, 12 slices/slab, 2 slice slab overlap, partial Fourier (PF) factor 3/4 and b=1000 s/mm².

*Experiment 1 Simulation Evaluation*: To validate the proposed sampling and reconstruction, we acquired single-slab data (12 slices) from one subject at 1.22 mm (Table 1. 1A) and 1.09 mm (Table 1. 1B). Fully sampled reference data were reconstructed from three scans with 0, 1, and 2Δky shift. The scanner used PF=3/4. Realistic phase-corrupted multi-shot data were simulated by multiplying the reference data with navigator phase maps followed by applying the shot sampling masks. We evaluated conventional and Self-nav CAIPI sampling (Fig. 2d, e). We also simulated Self-nav CAIPI sampling with and without the use of a non-blipped kz=0 shot and investigated the impact of including the magnitude constraint $\left|\left|\hat{x} - \rho' \Phi^{k-1}\right|\right|_2^2$ in Eq. 18.

*Experiment 2 SNR Evaluation with Matched TR*: To quantitatively evaluate SNR, we acquired data from three subjects using conventional (Table 1. 2A) and Self-nav CAIPI sampling (Table 1. 2B) with matched TE and TR. To evaluate the accuracy of phase map estimation, navigators were also acquired. Diffusion weighting along the readout, phase, and slice directions was evaluated, each with twelve repetitions.

*Experiment 3 SNR Efficiency Evaluation with Optimal TR*: To quantitatively compare the SNR efficiency, we acquired data in four subjects with conventional (with navigators, Table 1. 3A) and Self-nav CAIPI (without navigators, Table 1. 3B) sampling, with different TR optimized for each scan. Diffusion weighting was along readout with twelve repetitions.

*Experiment 4 DTI Comparison*: To compare DTI results, we acquired data in two subjects with 16 diffusion encoding directions with the conventional sampling (with navigators, Table 1. 4A) and the Self-nav CAIPI sampling (without navigators, Table 1. 4B). The two b=0 images were acquired along opposite phase encoding directions (1 blip-up and 1 blip-down) using conventional rectangular sampling.

*Experiment 5 High-b-value Protocol*: To demonstrate higher b-values, we acquired data from one subject with the conventional sampling (Table 1. 5A, C) and the Self-nav CAIPI sampling (Table 1. 5B, D) (both with navigators) at b=2000 s/mm² (Table 1. 5A, B) and b=3000 s/mm² (Table 1. 5C, D).

*Experiment 6 Tractography Protocol*: To demonstrate tractography analysis, we acquired data in one subject with Self-nav CAIPI sampling and 48 diffusion directions along with 6 b=0 images (3 blip-up and 3 blip-down) (Table 1. 6).

### B. Reconstruction details

The SPIRiT kernel in Eq. 13 was trained using gradient echo coil calibration data. The image reconstruction was conducted in MATLAB 2021a (Mathworks, Natick, MA, USA). The 32-channel data were compressed to 8 channels [25].

For the 3D data reconstruction in Eq. 20, the k-space data were first Fourier transformed along kx followed by reconstruction performed for each ky-kz plane using the SPIRiT kernel $G_{slab}$. The reconstructed 2D images were concatenated along the readout direction (x) to form the whole image volume. The kernel sizes for $G_0$ and $G_{slab}$ were both set to 5×5. The kernel size $w_H$ for constructing $H(\hat{x})$ was set to 10 (i.e., 10×10 kernel). The hyperparameters $\lambda_1, \lambda_3, \lambda_4, \beta$ were empirically selected as 1, 1e-4, 1e-4, 10, respectively. The SVD thresholding for solving Eq. 16 was performed by retaining the largest N singular values and their corresponding vectors, where $N = w_H^2$ (i.e., N=100 in this study). The ADMM iteration number was set to 50. The CG iteration numbers for solving Eqs. 18 and 20 were both set to 30.

The 2D navigator images were reconstructed with 2D GRAPPA and filtered by a k-space Hamming window of size 24×24. The phase images were extracted and used as an estimation of motion induced phase errors. The SLR estimated $\phi_d$ were also filtered using the same Hamming window for the reconstruction in Eq. 20.

TABLE I
ACQUISITION PARAMETERS

| Exp.# | Voxel size (mm$^3$) | TE1 (imaging) / TE2 (navigator) / TR(ms)[a] | b-value (s/mm$^2$) | #b=0 /DWI[b] | Sampling[c] | Navigator | T$_{acq}$ (min:sec) |
|---|---|---|---|---|---|---|---|
| 1A | 1.22 | 82/142/2000 | 1000 | 0/3 | Conventional | Yes | 1:12 |
| 1B | 1.09 | 89/157/3500 | 1000 | 0/3 | Conventional | Yes | 2:06 |
| 2A | 1.09 | 65/133/3500 | 1000 | 0/36 | Conventional | Yes | 25:12 |
| 2B | 1.09 | 65/133/3500 | 1000 | 0/36 | Self-nav CAIPI | Yes | 25:12 |
| 3A | 1.09 | 65/133/3500 | 1000 | 0/12 | Conventional | Yes | 8:24 |
| 3B | 1.09 | 64/-/2400 | 1000 | 0/12 | Self-nav CAIPI | No | 5:46 |
| 4A | 1.09 | 65/133/3500 | 1000 | 2/16 | Conventional | Yes | 12:36 |
| 4B | 1.09 | 64/-/2400 | 1000 | 2/16 | Self-nav CAIPI | No | 8:38 |
| 5A | 1.09 | 72/140/3500 | 2000 | 0/1 | Conventional | Yes | 0:42 |
| 5B | 1.09 | 72/140/3500 | 2000 | 0/1 | Self-nav CAIPI | Yes | 0:42 |
| 5C | 1.09 | 78/146/3500 | 3000 | 0/1 | Conventional | Yes | 0:42 |
| 5D | 1.09 | 78/146/3500 | 3000 | 0/1 | Self-nav CAIPI | Yes | 0:42 |
| 6 | 1.09 | 64/-/2400 | 1000 | 6/48 | Self-nav CAIPI | No | 25:55 |

a. 3/4 partial Fourier along phase encoding direction was applied for Exp. 2-6. The TRs of Exp. 2-6 were the shortest achievable TRs due to SAR restriction. b. The number of DWIs refers to the number of scans with different encoding for each experiment. In Exp. 1, three $\Delta k_y$ shift was acquired with the same diffusion encoding. In Exp. 2, 12 repetitions along three orthogonal diffusion encoding directions were obtained for SNR evaluation. In Exp. 3, 12 repetitions were acquired with a single diffusion encoding direction along readout. c. The Self-nav CAIPI samplings for 1.22 mm and 1.09 mm are illustrated in Fig. 2d and e respectively. All b=0 data were acquired with the conventional rectangular sampling.

## C. Image analysis

Image post-processing was conducted using the FMRIB Software Library (FSL) [26] unless indicated otherwise. For Exp. 1, the reconstructed images from different methods were evaluated by calculating normalized root mean squared error (NRMSE) with the fully sampled ground truth data. For Exp. 2, 3, and 5, the whole-brain diffusion-weighed data were obtained by directly combining multiple slabs and averaging the overlapped slices. To mitigate the impact of subject motion on SNR evaluation, all multi-repetition data were co-registered to the first repetition using "flirt" [27] with 6 degrees of freedom and trilinear interpolation. An interim mean image was computed by averaging all co-registered data, and the two volumes with the highest NRMSE with respect to this interim mean image were discarded to reduce motion contamination. The remaining ten-repetition data, denoted as $S_{10}$, were used for SNR calculation. The voxel-wise SNR map was calculated as mean($S_{10}$)./std($S_{10}$) (./ denotes the element-wise division). The SNR efficiency (i.e., SNR per unit time) was calculated as $SNR/\sqrt{TR}$. The mean SNR and SNR efficiency were then calculated by taking the mean of the SNR and SNR efficiency maps within a brain mask, respectively.

For DTI and tractography experiments (Exp. 4 and 6), slab combination and correction for slab saturation artifacts were performed for the diffusion data using nonlinear inversion of slab profile encoding (NPEN) [28]. Whole-brain images were corrected for Gibbs ringing [29]. A motion field map was estimated using blip-reversed b=0 image volumes using "topup" [30], which was then input to "eddy" [31] along with all diffusion data to correct for off-resonance distortions, eddy current effects, and subject motion. The diffusion analyses were conducted in the native diffusion space. For Exp. 4, the diffusion tensor model fitting was performed using "dtifit". For Exp. 6, white matter tractography was performed using "autoPtx" [32] including probabilistic model fit "bedpostx" and probabilistic tractography "probtrackx".

## IV. RESULTS

Fig. 4 demonstrates the image reconstruction results from the simulation data. For conventional rectangular sampling, the reconstructed images without phase error correction suffer from severe artifacts and exhibit high NRMSE values (Fig. 4, iii). Employing navigator phase maps leads to a significant enhancement in image quality but requires additional scan time (Fig. 4, ii). In contrast, our proposed self-navigated approach avoids the necessity for navigators while achieving comparable efficacy in phase error correction with conventional navigated acquisition (Fig. 4, v). The optimized Self-nav CAIPI sampling produces similar image quality with navigator phase (Fig. 4, iv) and SLR-estimated phase (Fig. 4, v), demonstrating the robust phase variance estimation of the self-navigated approach. The residual maps exhibit random-noise-like error (Fig. 4b, d, iv, v) without anatomical structure, likely due to the pseudo-random sampling patterns.

The image reconstruction performance of the proposed method depends on the accuracy of the motion-induced phase variation maps. As show in Fig. 5, using the proposed SPIRiT SLR reconstruction, the estimated phase maps are highly similar to the ground truth phase maps with consistent spatial patterns for all shots, which confirms that our proposed self-navigated method is able to provide accurate estimation of motion induced phase errors.

Fig. 6 illustrates the impact of including the shot traversing the kz=0 plane. When the kz=0 traversing shot is replaced with another CAIPI shot obtained from greedy search in Eq. 11, the SLR reconstruction produces a phase map with lower accuracy (Fig. 6b, iv) with a slower convergence (Fig. 6c). The lower phase accuracy affects the multi-shot reconstruction performance, leading to a higher reconstruction error (Fig. 6a, iv). Having the magnitude constraint $\left\| \hat{x} - \rho'\Phi^{k-1} \right\|_2^2$ in Eq. 18 is beneficial for further accelerating the convergence (Fig. 6c).

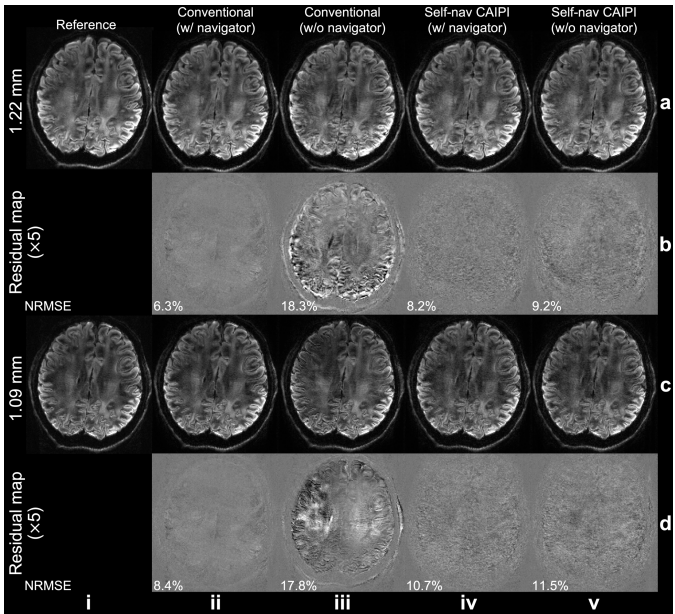

Fig. 4. Image reconstruction results from the simulation experiment. Retrospectively under-sampled images from Simulation Evaluation (Exp. 1) are shown, reconstructed using the following methods: conventional sampling with ground truth phase maps (ii), conventional sampling without phase error correction (iii), self-navigated CAIPI sampling with ground truth phase maps (iv), and self-navigated CAIPI sampling with structured low-rank estimated phase maps (v), and at 1.22 mm (a) and 1.09 mm (c) isotropic resolutions. The residual maps (b and d) with respect to the fully sampled ground truth data (i) are also demonstrated. The normalized root mean squared errors (NRMSE) of the entire slab are provided to quantify the image similarity.

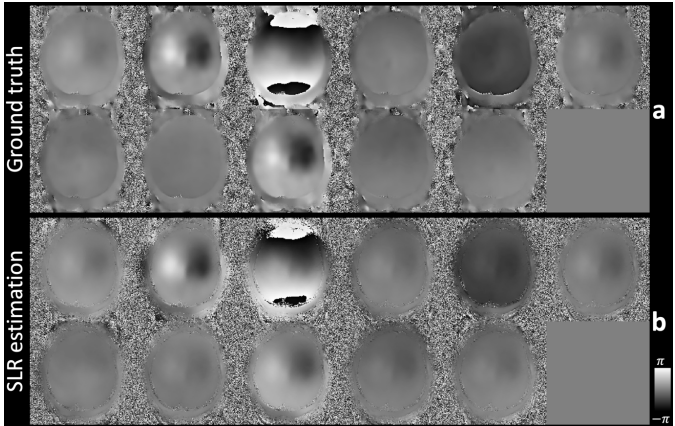

Fig. 5. Phase map estimation from the simulation experiment. The ground truth phase maps (a) and structured low-rank (SLR) estimated phase maps (without filtering) from the proposed method (b) are shown for different shots of the simulated data from Simulation Evaluation (Exp. 1) at 1.22 mm isotropic resolutions. The phase of the shot traversing the entire kz=0 is subtracted from all shots to eliminate the phase offsets, effectively demonstrating the motion induced phase variations across shots (11 shots demonstrated).

The proposed method works well on prospectively acquired in-vivo data (Fig. 7). While the conventional method exhibits significant image artifacts when navigators are not acquired (Fig. 7a, ii, iii), our proposed self-navigated method achieves similar image reconstruction results using either navigator phase maps or SLR estimated phase maps (Fig. 7b). Notably, the navigator acquired phase map and estimated phase map exhibit high consistency (Fig. 7c).

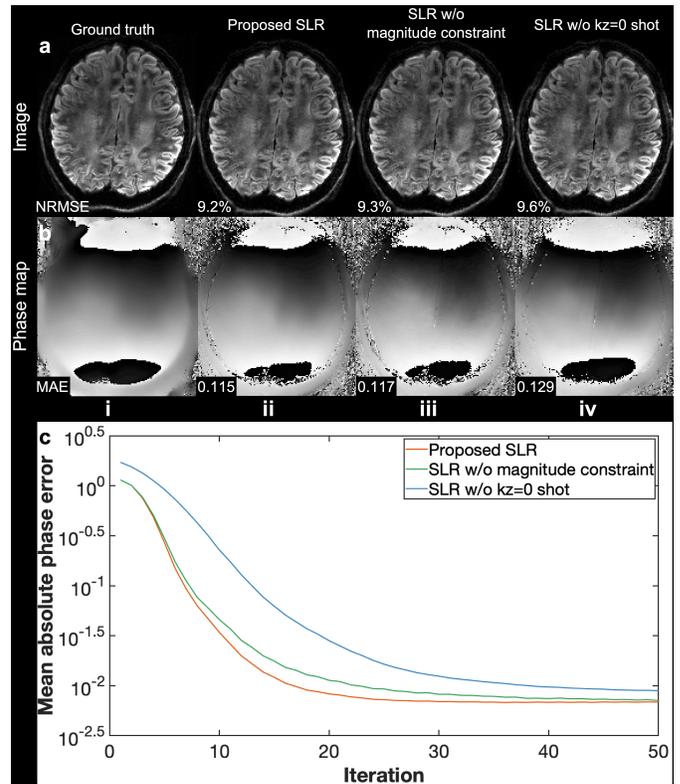

Fig. 6. Impact of including the shot traversing the kz=0 plane. Retrospectively under-sampled images (a) and 2D phase maps (without filtering) (b) of one shot from Simulation Evaluation (Exp. 1) are shown. Reconstructions using the proposed structured low-rank (SLR) (ii), SLR without the magnitude constraint $\left\| \tilde{x} - \rho' \Phi^{k-1} \right\|_2^2$ in Eq. 18 (iii) and SLR without the kz=0 traversing shot (substituted with a greedy-searched CAIPI shot) (iv) are compared. Image resolution is 1.22 mm isotropic. The normalized root mean squared errors (NRMSE) of the entire image slab and the mean absolute error (MAE) of phase maps of all shots within the brain mask compared to the ground truth (i) are provided to quantify the image similarity and phase estimation accuracy. The phase MAE plot of different SLR strategies at different ADMM iterations is also shown to demonstrate the convergence of the optimization (c).

Our proposed method exhibits robust performance along different diffusion encoding directions (Fig. 8). When acquired with the same TR, the self-navigated method (Fig. 8d) produces SNR values comparable to the conventional navigated method (Fig. 8a), even along the slice selection diffusion encoding direction (Fig. 8, iii) where motion-induced phase variance is most significant [7]. Notably, the SNR values obtained from our self-navigated method are substantially higher than those from conventional sampling without phase error correction and even marginally higher than those from navigated Self-nav CAIPI sampling. This could be attributed to the higher resolution of SLR reconstructed phase maps compared to navigators (64 phase encoding lines acquired) which may fail to capture high-frequency phase changes in certain cases (i.e., larger motions). The group-level SNR values for three subjects (Exp. 2, Table II) are consistent with the findings depicted in Fig. 8.

Removing the navigator acquisition can shorten TR, leading to an improved SNR efficiency. With the TR reduction, the SNR obtained from the self-navigated method is slightly lower compared to the conventional navigated method (Fig. 9, i), as would be expected due to the reduced T1 recovery. For the four subjects scanned in Exp. 3, the group-level SNR values (mean

± std) are $11.70 \pm 0.74$ and $11.18 \pm 0.41$ for the conventional and proposed method, respectively. In terms of SNR efficiency, the group-level results for the conventional and proposed methods are $6.25 \pm 0.39$ and $7.22 \pm 0.26$, respectively. The self-navigated method improves SNR efficiency by 15.5% compared to the conventional method.

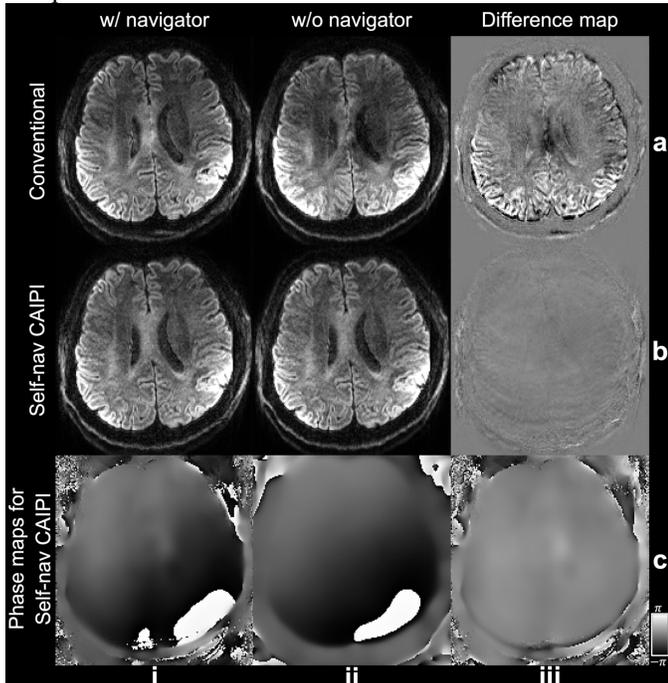

Fig. 7. Reconstruction of prospectively acquired data. In-vivo diffusion weighted (diffusion encoding along the slice selection direction) images at 1.09 mm isotropic resolution of a representative subject from Exp. 2 are shown. Reconstruction with navigators (i) and without navigators (ii) using conventional rectangular sampling (a) and Self-nav CAIPI sampling (b). The difference maps (iii) between i and ii are also presented. Additionally, the navigator acquired phase map and structured low-rank reconstructed phase map of a representative shot of Self-nav CAIPI sampling, along with their difference, are displayed (c). For conventional sampling, "without navigator" indicates the absence of phase error correction. For self-navigated CAIPI sampling, "without navigator" refers to the utilization of structured low-rank estimated phase for reconstruction.

The self-navigated method enables high-quality DTI within a shorter scan time (Fig. 9, ii). Our method obtained the DTI maps within 8.6 minutes (Fig. 9b, ii), with comparable SNR to the conventional method requiring 12.6 minutes (Fig. 9a, ii). By eliminating the navigator acquisition, we save 31.7% of the total scan time without noticeable loss in image quality.

Fig. 10 demonstrates the performance of our method at higher b-values, a scenario where the diffusion data have lower SNR and are more sensitive to motion-induced phase errors. The Self-nav CAIPI sampling with SLR estimated phase maps (Fig. 10, iii) produces similar results to conventional rectangular sampling (Fig. 10, i) and Self-nav CAIPI sampling with navigator phase maps (Fig. 10, ii). The SLR estimated phase maps (Fig. 10, v) are highly similar to the navigator phase maps (Fig. 10, iv), even at high b-values (e.g., b=3000 s/mm²). This result demonstrates the robustness of the proposed sampling and reconstruction method at high b-values.

Fig. 11 demonstrates whole-brain tractography results using the proposed self-navigated method, with 14 tracts illustrated in coronal and sagittal views. The data acquired with the proposed method support the delineation of major fiber bundles.

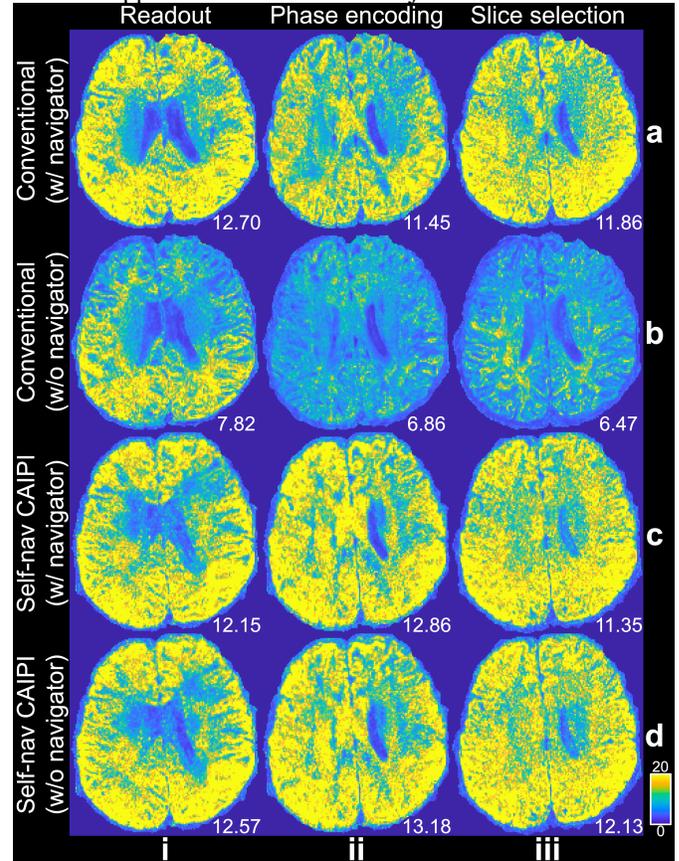

Fig. 8. SNR maps of different diffusion encoding directions. The SNR maps for diffusion-weighted images (1.09 mm isotropic resolution) along readout (i), phase encoding (ii), and slice selection (iii) diffusion encoding directions from conventional and Self-nav CAIPI acquisition with the same TR of a representative subject from Exp. 2 are demonstrated. Four acquisition and reconstruction methods are compared: (a) conventional rectangular sampling and reconstruction with navigator phase maps. (b) conventional rectangular sampling and reconstruction without phase error correction. (c) Self-nav CAIPI acquisition and reconstruction with navigator phase maps. (d) Self-nav CAIPI acquisition and reconstruction with structured low-rank estimated phase maps (d). The mean SNR calculated within a brain mask for this subject is listed.

TABLE II
GROUP-LEVEL SNR VALUES ALONG DIFFERENT DIFFUSION ENCODING DIRECTIONS

| | Readout | Phase encoding | Slice selection |
|---|---|---|---|
| Convectional (w/ navigator) | 11.66±1.23 | 11.33±0.37 | **11.65±0.38** |
| Conventional (w/o navigator) | 7.37±0.53 | 6.91±0.13 | 6.14±0.33 |
| Self-nav CAIPI (w/ navigator) | 11.46±0.72 | 12.46±0.45 | 10.57±0.73 |
| Self-nav CAIPI (w/o navigator) | **11.95±0.60** | **13.04±0.16** | 11.43±0.81 |

For conventional sampling, "without navigator" indicates the absence of phase error correction. For self-navigated CAIPI sampling, "without navigator" refers to structured low-rank estimated phase correction. The highest SNR for each diffusion direction is marked in bold.

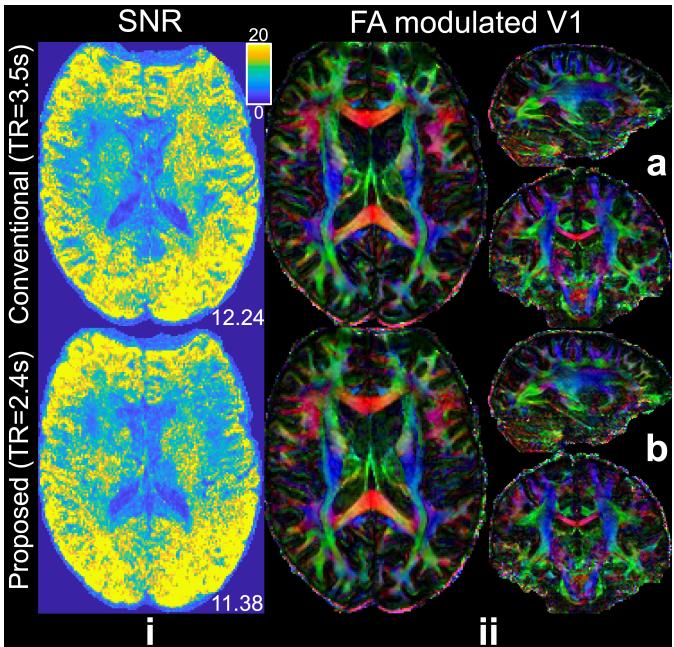

Fig. 9. SNR efficiency and DTI comparisons. The comparison of SNR maps (Exp. 3, i) and DTI results (Exp. 4, ii, 1.09 mm isotropic resolution) of a representative subject is presented. The SNR maps and DTI results are obtained using the conventional navigated method (a) and the proposed self-navigated method (b). The whole-brain SNR values (calculated within a brain mask) for this subject are listed. Notably, the proposed method has a shorter TR by eliminating navigator acquisition.

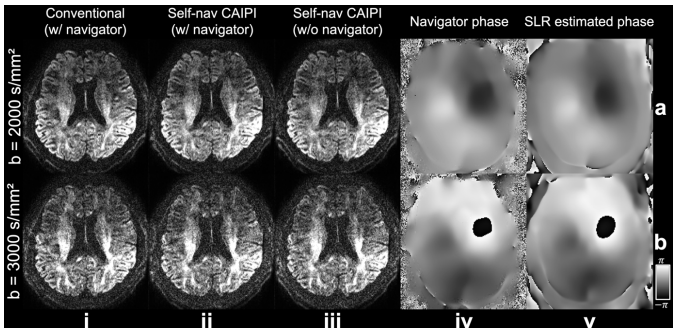

Fig. 10. Results at high b-values. Diffusion-weighted images (1.09 mm isotropic resolution, diffusion encoding along readout) reconstructed using conventional sampling with navigator (i), Self-nav CAIPI sampling with navigator (ii), and Self-nav CAIPI sampling with structured low-rank estimated phase maps (iii) at b=2000 s/mm² (a) and 3000 s/mm² (b) from Exp. 5 are shown. The phase maps of one representative shot for Self-nav CAIPI sampling from the navigator (iv) and the proposed structured low-rank (SLR) reconstruction (v) are also demonstrated.

## V. DISCUSSION

In this work, we propose a novel acquisition and reconstruction framework that extends 2D SLR-based phase correction [11, 12, 14] to 3D multi-slab diffusion MRI to eliminate the need for acquiring navigators. Each shot of our Self-nav CAIPI sampling intersects with the central kz=0 plane, providing self-navigation points. The sampling is optimized for overlapping, k-space gaps, and self-navigation performance using a greedy search algorithm. Self-nav CAIPI sampling allows us to leverage a SLR reconstruction method that exploits the redundancy across shots and coils to obtain a 2D phase map from self-navigation points for each shot. One shot traverses the entire kz=0 plane to provide accurate magnitude information, accelerating the convergence and improving the robustness of the reconstruction. In-vivo experiments from seven subjects validate the efficacy and robustness of our proposed method, which saves 31.7% of scan time by eliminating the navigator acquisition and achieves 15.5% higher SNR efficiency compared to conventional navigated 3D multi-slab imaging.

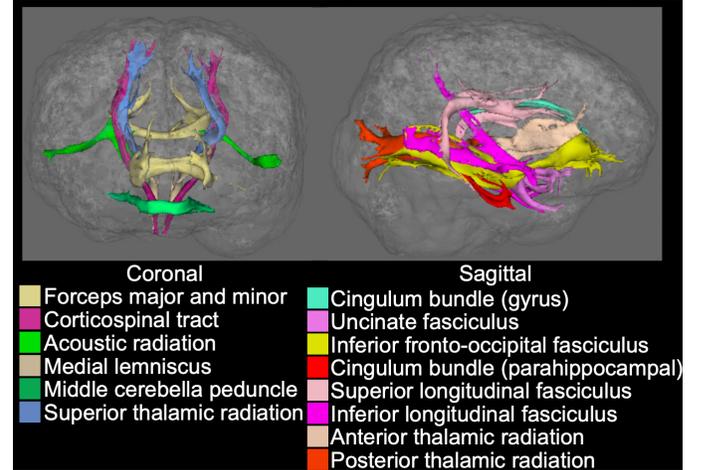

Fig. 11. Tractography results. The binarized tractography (threshold: 0.01) results obtained from 48-direction diffusion data at 1.09 mm isotropic resolution using our proposed method from Exp. 6 are presented. Fourteen tracts are illustrated in coronal and sagittal views.

Our sampling design hinges on two critical considerations: optimizing for the multi-shot reconstruction and self-navigation performance (i.e., the reconstruction of kz=0 plane). We developed an optimization framework to determine the optimal sampling pattern using a shot-by-shot greedy search. Given the substantial parameter space (with $s_{k_z} \in [0,5]$, $s_{k_y} \in [0,2]$, $s_p \in [0,11]$), conducting a global search across 11 shots would involve evaluating $(6 \times 3 \times 12)^{11} \approx 4.78 \times 10^{25}$ sampling patterns. Our greedy search approach significantly reduces the search space to $6 \times 3 \times 12 \times 11 = 2376$ while achieving satisfactory reconstruction performance. Furthermore, the uniform and random samplings may benefit compressed sensing reconstruction methods [19], indicating that adding tailored constraints in Eq. 20 may further enhance performance.

Due to the limited number of self-navigation points, the acceleration factor for kz=0 reconstruction is extremely high (R=18 or 36). Our SLR constraint leverages the magnitude similarity across shots and redundant information across coils. The shot traversing kz=0 provides faster convergence (Fig. 6) and reduces the unknowns from the complex image to only the phase image. The optimized self-navigation performance (i.e., $d_i$ in Eq. 11) ensures that each shot contains sufficient low-frequency information to produce a reliable phase map. These attributes enable robust reconstruction of 2D phase maps from limited data, even with low-SNR (e.g., high b-values, Fig. 10).

The SLR estimated phase maps demonstrate slightly better reconstruction performance than navigator-acquired phase maps in prospective in-vivo experiments (Fig. 8, Table 2). This may be because the reconstructed phase maps have the same resolution as the imaging data, whereas navigators are typically low resolution. In Exp. 1, the structured low-rank assisted reconstruction exhibit slightly higher NRMSE compared to navigated reconstruction, presumably because here the navigator phase maps are ground truth phase maps that were retrospectively added to phase error-free data; in practice navigators contain some measurement error.

The inclusion of both shot and coil dimensions contributes to the substantial size of the Hankel matrix $H(\hat{x})$ (Fig. 3), making the current reconstruction time-consuming. Each iteration of the ADMM takes approximately 8 minutes on a 2.9 GHz Quad-Core Intel Core i7 CPU. The inclusion of the kz=0 traversing shot and the magnitude constraint $\left\| \hat{x} - \rho' \Phi^{k-1} \right\|_2^2$ may help reduce the number of iterations without compromising the reconstructed phase map quality (Fig. 6c). Model-based deep learning reconstruction approaches [14] may also offer a promising avenue for accelerating the reconstruction process.

## VI.  CONCLUSION

We present a novel acquisition and reconstruction framework that eliminates the requirement for navigator acquisition in high-resolution 3D multi-slab diffusion MRI. It effectively shortens the TR and reduces SAR, enhancing scan efficiency and safety, and permitting the use of better RF pulses. The effectiveness of our method is evident in high-fidelity fast DTI and tractography and can be explored in more applications.

## REFERENCES

[1]  K. L. Miller *et al.*, "Diffusion imaging of whole, post-mortem human brains on a clinical MRI scanner," *Neuroimage,* vol. 57, no. 1, pp. 167-181, 2011.

[2]  J. A. McNab *et al.*, "Surface based analysis of diffusion orientation for identifying architectonic domains in the in vivo human cortex," *Neuroimage,* vol. 69, pp. 87-100, 2013.

[3]  R. Frost, K. L. Miller, R. H. Tijssen, D. A. Porter, and P. Jezzard, "3D Multi‑slab diffusion‑weighted readout‑segmented EPI with real‑time cardiac‑reordered k‑space acquisition," *Magnetic resonance in medicine,* vol. 72, no. 6, pp. 1565-1579, 2014.

[4]  W. Wu *et al.*, "High‑resolution diffusion MRI at 7T using a three‑dimensional multi‑slab acquisition," *NeuroImage,* vol. 143, pp. 1-14, 2016.

[5]  M. Engström and S. Skare, "Diffusion‑weighted 3D multislab echo planar imaging for high signal‑to‑noise ratio efficiency and isotropic image resolution," *Magnetic resonance in medicine,* vol. 70, no. 6, pp. 1507-1514, 2013.

[6]  Z. Li *et al.*, "Sampling strategies and integrated reconstruction for reducing distortion and boundary slice aliasing in high‑resolution 3D diffusion MRI," *Magnetic Resonance in Medicine,* 2023.

[7]  K. L. Miller and J. M. Pauly, "Nonlinear phase correction for navigated diffusion imaging," *Magnetic Resonance in Medicine: An Official Journal of the International Society for Magnetic Resonance in Medicine,* vol. 50, no. 2, pp. 343-353, 2003.

[8]  K. Butts, A. de Crespigny, J. M. Pauly, and M. Moseley, "Diffusion‑weighted interleaved echo‑planar imaging with a pair of orthogonal navigator echoes," *Magnetic resonance in medicine,* vol. 35, no. 5, pp. 763-770, 1996.

[9]  S. Moeller *et al.*, "Self‑navigation for 3D multishot EPI with data‑reference," *Magnetic resonance in medicine,* vol. 84, no. 4, pp. 1747-1762, 2020.

[10]  N.-k. Chen, A. Guidon, H.-C. Chang, and A. W. Song, "A robust multi-shot scan strategy for high-resolution diffusion weighted MRI enabled by multiplexed sensitivity-encoding (MUSE)," *Neuroimage,* vol. 72, pp. 41-47, 2013.

[11]  M. Mani, M. Jacob, D. Kelley, and V. Magnotta, "Multi‑shot sensitivity‑encoded diffusion data recovery using structured low‑rank matrix completion (MUSSELS)," *Magnetic resonance in medicine,* vol. 78, no. 2, pp. 494-507, 2017.

[12]  M. Mani, H. K. Aggarwal, V. Magnotta, and M. Jacob, "Improved MUSSELS reconstruction for high‑resolution multi‑shot diffusion weighted imaging," *Magnetic resonance in medicine,* vol. 83, no. 6, pp. 2253-2263, 2020.

[13]  I. Markovsky, "Structured low-rank approximation and its applications," *Automatica,* vol. 44, no. 4, pp. 891-909, 2008.

[14]  H. K. Aggarwal, M. P. Mani, and M. Jacob, "MoDL-MUSSELS: model-based deep learning for multishot sensitivity-encoded diffusion MRI," *IEEE transactions on medical imaging,* vol. 39, no. 4, pp. 1268-1277, 2019.

[15]  C. Liao *et al.*, "Distortion‑free, high‑isotropic‑resolution diffusion MRI with gSlider BUDA‑EPI and multicoil dynamic B0 shimming," *Magnetic resonance in medicine,* vol. 86, no. 2, pp. 791-803, 2021.

[16]  C. Liao *et al.*, "High-fidelity mesoscale in-vivo diffusion MRI through gSlider-BUDA and circular EPI with S-LORAKS reconstruction," *NeuroImage,* vol. 275, p. 120168, 2023.

[17]  F. A. Breuer *et al.*, "Controlled aliasing in volumetric parallel imaging (2D CAIPIRINHA)," *Magnetic Resonance in Medicine: An Official Journal of the International Society for Magnetic Resonance in Medicine,* vol. 55, no. 3, pp. 549-556, 2006.

[18]  M. Lustig and J. M. Pauly, "SPIRiT: iterative self‑consistent parallel imaging reconstruction from arbitrary k‑space," *Magnetic resonance in medicine,* vol. 64, no. 2, pp. 457-471, 2010.

[19]  M. Seeger, H. Nickisch, R. Pohmann, and B. Schölkopf, "Optimization of k‑space trajectories for compressed sensing by Bayesian experimental design," *Magnetic Resonance in Medicine: An Official Journal of the International Society for Magnetic Resonance in Medicine,* vol. 63, no. 1, pp. 116-126, 2010.

[20]  M. Murphy, M. Alley, J. Demmel, K. Keutzer, S. Vasanawala, and M. Lustig, "Fast $\ell_1$‑S-SPIRiT compressed sensing parallel imaging MRI: scalable parallel implementation and clinically feasible runtime," *IEEE transactions on medical imaging,* vol. 31, no. 6, pp. 1250-1262, 2012.

[21]  P. J. Shin *et al.*, "Calibrationless parallel imaging reconstruction based on structured low‑rank matrix completion," *Magnetic resonance in medicine,* vol. 72, no. 4, pp. 959-970, 2014.

[22]  M. A. Griswold *et al.*, "Generalized autocalibrating partially parallel acquisitions (GRAPPA)," *Magnetic Resonance in Medicine: An Official Journal of the International Society for Magnetic Resonance in Medicine,* vol. 47, no. 6, pp. 1202-1210, 2002.

[23]  S. Boyd, N. Parikh, E. Chu, B. Peleato, and J. Eckstein, "Distributed optimization and statistical learning via the alternating direction method of multipliers," *Foundations and Trends® in Machine learning,* vol. 3, no. 1, pp. 1-122, 2011.

[24]  X. Chen, W. Wu, and M. Chiew, "Improving robustness of 3D multi-shot EPI by structured low-rank reconstruction of segmented CAIPI sampling for fMRI at 7T," *NeuroImage,* vol. 267, p. 119827, 2023.

[25]  T. Zhang, J. M. Pauly, S. S. Vasanawala, and M. Lustig, "Coil compression for accelerated imaging with Cartesian sampling," *Magnetic resonance in medicine,* vol. 69, no. 2, pp. 571-582, 2013.

[26]  M. Jenkinson, C. F. Beckmann, T. E. Behrens, M. W. Woolrich, and S. M. Smith, "Fsl," *Neuroimage,* vol. 62, no. 2, pp. 782-790, 2012.

[27]  M. Jenkinson and S. Smith, "A global optimisation method for robust affine registration of brain images," *Medical image analysis,* vol. 5, no. 2, pp. 143-156, 2001.

[28]  W. Wu, P. J. Koopmans, R. Frost, and K. L. Miller, "Reducing slab boundary artifacts in three‑dimensional multislab diffusion MRI using nonlinear inversion for slab profile encoding (NPEN)," *Magnetic resonance in medicine,* vol. 76, no. 4, pp. 1183-1195, 2016.

[29]  T. Bautista, J. O'Muircheartaigh, J. Hajnal, and T. J-Donald, "Removal of Gibbs ringing artefacts for 3D acquisitions using subvoxel shifts," in *Proc Int Soc Magn Reson Med,* 2021, vol. 29, p. 3535.

[30]  J. L. Andersson, S. Skare, and J. Ashburner, "How to correct susceptibility distortions in spin-echo echo-planar images: application to diffusion tensor imaging," *Neuroimage,* vol. 20, no. 2, pp. 870-888, 2003.

[31]  J. L. Andersson and S. N. Sotiropoulos, "An integrated approach to correction for off-resonance effects and subject movement in diffusion MR imaging," *Neuroimage,* vol. 125, pp. 1063-1078, 2016.

[32]  M. De Groot *et al.*, "Improving alignment in tract-based spatial statistics: evaluation and optimization of image registration," *Neuroimage,* vol. 76, pp. 400-411, 2013.